\documentstyle[preprint,aps]{revtex}
\begin{document}
\draft
\begin{title}
{\bf SOLITONS IN POLYMERIC CHAINS\\
      WITH PERIODIC INTERACTIONS}
\end{title}

\author{D. Bazeia, R. F. Ribeiro, and E. Ventura}

\address{Departamento de F\'\i sica, Universidade Federal da Para\'\i ba\\
Caixa Postal 5008, 58051-970 Jo\~ao Pessoa, Para\'\i ba,  Brazil}

\author{M. T. Thomaz}

\address{Instituto de F\'\i sica, Universidade Federal Fluminense\\
R. Gal. Milton Tavares de Souza s/ n.$\!\!^{\rm o}$ \\
24210-340 Niter\'oi, Rio de Janeiro, Brazil}

\maketitle

\begin{abstract}
In this paper we follow the lines of recent works to investigate systems
of two coupled real scalar fields defined by potentials that describe periodic 
interactions between the scalar fields. We work with polymeric chains
containing periodic interactions between the coupled fields, and we
investigate the topological sectors to obtain explicit soliton solutions
and their corresponding energy. In particular, we offer an example that
considers deoxyribonucleic acid (DNA) as a system of coupled fields,
and we present the main steps to describe DNA as a polymeric chain belonging
to the class of systems of two coupled real scalar fields.
\end{abstract}
\vskip 1cm



\section{Introduction}

This paper deals with solitons in nonlinear systems of two coupled 
real scalar fields, described by periodic interactions between them 
in bidimensional spacetime. This is important in nonlinear science
in general, since  there are many systems which are described by periodic
interactions among their relevant degrees of freedom. The present work is a
continuation of former papers \cite{bds95,bsa95,brs95}, in which we have set
the basic ideas  we are now going to enlarge to include some specific
systems that are described by periodic interactions. Our aim
here is to investigate the presence of soliton solutions in systems belonging
to a general class of systems of coupled real scalar fields, and containing 
periodic interactions. Evidently, we believe that such systems can
engender interesting physical contents, and can be used to map relevant
degrees of freedom in quasi-one-dimensional periodic chains usually considered
to model actual systems in condensed matter, organic chemistry and biological
science. Within this context, in this work we use the term polymeric chains
to name these systems generically.

For simplicity, we consider  the case of two fields, namely
$\phi$ and $\chi$. Systems of this kind are usually described by 
a Lagrangian density that contains a potential $U=U(\phi,\chi)$,
in general a nonlinear function of the real scalar fields. It is 
this nonlinearity that enlarges the scope of the problem, since it 
can be mapped to many interesting systems in nonlinear science. 
In this paper, in particular, we shall investigate periodic interactions
that can be  of great interest in applications to nonlinear science due to
their topological properties, which we are also going to explore below. 
In this case, as an illustration we shall consider the deoxyribonucleic 
acid (DNA) as a system of coupled fields. In this first atempt to describe
DNA as a system of coupled fields belonging to the class already introduced
in \cite{brs95}, the main motivation is to shed some light on the way this
can be done. Since the basic motivation of the present work is to introduce
generic polymeric chains and to deal mainly with their mathematical
intricacies, we postpone to a future work a more specific investigation of
DNA as a system of two coupled real scalar fields. Here we recall another
recent work \cite{bnt97}, in which an alternate route to solitons in
hydrogen bonded chains is introduced.

The paper is organized as follows. In the next Sec.~{\ref{sec:system}} 
we briefly review the method of searching for soliton solutions we 
have already introduced in \cite{bds95,bsa95,brs95}. There we introduce 
a general Lagrangian density and set the conditions the potential has 
to obey, such that the static solution of the second-order equations
of motion are obtained by solving  first-order differential equations.
In Sec.~{\ref{sec:periodic}} we examine  examples of periodic systems.
These examples illustrate how to search for soliton solutions in systems
described by periodic interactions between two coupled scalar fields.
In Sec.~{\ref{sec:dna}} we consider DNA as a two-field system, and we show
how to describe it via the class of system here considered. In this case, we
start from the model proposed by Watson and Crick, in which DNA is
formed by two polynucleotide strands that constitute the double helix
held together by hydrogen bonds. We end the paper in
Sec.~{\ref{sec:comments}}, and there we present some comments and conclusions.

\section{A General Class of Systems}
\label{sec:system}

A general Lagrangian density describing a relativistic system of
two coupled real scalar fields in bidimensional spacetime is given by
\begin{equation}
\label{eq:model}
{\cal L}= \frac{1}{2} \partial_{\alpha}\phi\partial^{\alpha}\phi +
\frac{1}{2} \partial_{\alpha}\chi\partial^{\alpha}\chi - U(\phi,\chi),
\end{equation}
where $U=U(\phi,\chi)$ is the potential, which specifies the
particular system one is interested in. Our notation is usual:
we are using natural units, in which $\hbar=c=1$, and the metric
tensor $g^{\alpha\beta}$ is diagonal, with $g^{00}=-g^{11}=1$. 
In this case the fields are dimensionless, and time $(t)$ and space 
coordinate $(x)$ have dimension inverse of energy. The potential will
be specified with two kinds of parameters: $(a,b,...)$ as
real and positive dimensionless parameters, and $(\lambda,\mu,...)$
as real parameters having dimension of energy.

The above system leads to the following set of equations of motion
\begin{equation}
\label{eq:motiont1}
\frac{\partial^2\phi}{\partial t^2}-\frac{\partial^2\phi}{\partial x^2} 
+ \frac{\partial U}{\partial\phi}=0,
\end{equation}
and
\begin{equation}
\label{eq:motiont2}
\frac{\partial^2\chi}{\partial t^2}-\frac{\partial^2\chi}{\partial x^2} 
+ \frac{\partial U}{\partial\chi}=0.
\end{equation}
In the standard way \cite{raj82} of searching for soliton 
solutions one consider static field configurations, and so 
$\phi=\phi(x)$ and $\chi=\chi(x)$. In this case,
the equations of motion become
\begin{equation}
\label{eq:motionx1}
\frac{d^2\phi}{dx^2} = \frac{\partial U}{\partial\phi},\label{4}
\end{equation}
and
\begin{equation}
\label{eq:motionx2}
\frac{d^2\chi}{dx^2} = \frac{\partial U}{\partial\chi}.\label{5}
\end{equation}

As it was already stressed in \cite{brs95}, let us now consider potentials
that can be written as
\begin{equation}
\label{eq:poten}
U(\phi,\chi)=\frac{1}{2}\left(\frac{\partial H}{\partial\phi}\right)^2 +
\frac{1}{2}\left(\frac{\partial H}{\partial\chi}\right)^2,
\end{equation}
where the functions $H=H(\phi,\chi)$ is a smooth but otherwise
arbitrary functions of the fields $\phi$ and $\chi$. Here the
equations of motion describing the static field configurations become
\begin{equation}
\label{eq:motionxs1}
\frac{d^2\phi}{dx^2} = H_{\phi}H_{\phi\phi}+H_{\chi}H_{\phi\chi},
\end{equation}
and
\begin{equation}
\label{eq:motionxs2}
\frac{d^2\chi}{dx^2} = H_{\phi}H_{\phi\chi}+H_{\chi}H_{\chi\chi},
\end{equation}
where we are using $H_{\phi}=\partial H/\partial\phi$ etc.

At first glance, it seems that we rewrote equations $(\ref{4})$ and
$(\ref{5})$ in a more complicated form, but this is not so, as it was
already shown \cite{bds95,bsa95,brs95}. In this case the energy of the
system can be cast to the form
\begin{equation}
\label{eq:energym}
E =\int_{-\infty}^{\infty} dx \; \frac{dH}{dx} =
 H(\phi(\infty),\chi(\infty))-
H(\phi(-\infty),\chi(-\infty)),
\end{equation}
and this is the minimum value for the energy, which is achieved when we
impose the conditions
\begin{equation}
\label{eq:foeq1}
\frac{d\phi}{dx}=H_{\phi},
\end{equation}
and
\begin{equation}
\label{eq:foeq2}
\frac{d\chi}{dx}=H_{\chi}.
\end{equation}

As we can easely see, solutions of the first-order equations (\ref{eq:foeq1}) 
and (\ref{eq:foeq2}) also satisfy  the second-order equations of motion
(\ref{eq:motionxs1}) and (\ref{eq:motionxs2}) since $H(\phi,\chi)$
is smooth. Therefore,  for the general system (\ref{eq:model}), when the
potential has the specific form (\ref{eq:poten}), the second-order
differential equations of motion (\ref{eq:motionxs1}) and
(\ref{eq:motionxs2}) can be replaced by the first-order differential
equations (\ref{eq:foeq1}) and (\ref{eq:foeq2}).

We omit details here, but it was already shown \cite{bsa95,brs95} that every
soliton solution this class of systems can comprise is classically or linearly
stable. Furthermore, in the above class of systems the energy corresponding
to static solutions is bounded from below, and gets to its minimum value
given by (\ref{eq:energym}). The function $H(\phi,\chi)$ can also be
used to define topological sectors, as shown in \cite{brs95}: The topological
charge, which is conserved, can be
cast to the form
\begin{equation}
Q_T=H(\phi(\infty),\chi(\infty))-H(\phi(-\infty),\chi(-\infty)),
\end{equation}
and in this case it is identified with the energy of the corresponding
static field configurations. However, since static field configurations
should go to vacuum states, asymptotically, to mantain the energy finite,
we see that the topological charge is nothing but the difference between
two values of $H(\phi,\chi)$, calculated at, say, $(\bar{\phi},\bar{\chi})$
and $(\bar{\phi}',\bar{\chi}')$, which represent two neighbour vacuum states
in the $(\phi,\chi)$ plane.

\section{Polymeric Chains with Periodic Interactions}
\label{sec:periodic}

The investigations introduced in the former Sect.~{\ref{sec:system}}
are done on general grounds, and can be generalized to the case of 
three or more fields straightforwardly. Here, however, our motivation is
to show how the general procedure we have already introduced works when
investigating specific systems. In former papers \cite{bds95,brs95}
we have already introduced several examples, in which we presented
explicit soliton solutions. There we have investigated systems
identified by potentials containing polynomial interactions between
two coupled fields. In this section we focus  our attention on
periodic interactions, and consider potentials described by the
sine and cosine functions.

As we have already seen, the basic function we have to take under 
consideration is $H(\phi,\chi)$, and this is the understanding
we shall now follow to investigate some explicit examples of systems
that contain periodic interactions between the two scalar fields.
The following polymeric chains are introduced with two main motivations:
Firstly, to deal with mathematical intricacies that naturally appear in
these systems and, secondly, as a preparation for possible applications to
actual systems.

\subsection{Polymeric Chain Number One}

As a first polymeric chain containing periodic interactions, let us choose
$H(\phi,\chi)$ in the form
\begin{equation}
\label{eq:permod2}
H_1(\phi,\chi)=\mu\, \cos(\phi)+\nu\,\cos(a\chi) + 
\lambda\, \cos(\phi)\, \cos(a\chi).
\end{equation}
Other forms similar to the above one can be obtained by just shifting
$\phi\to\phi+\pi/2 $ and/or $\chi\to\chi+\pi/2a$. In the present case,
however, the first-order equations obtained from $H_1(\phi,\chi)$ are
\begin{equation}
\label{eq:mod2a}
\frac{d\phi}{dx}+ \sin(\phi)[\mu+\lambda \cos(a\chi)]=0,
\end{equation}
and
\begin{equation}
\label{eq:mod2b}
\frac{d\chi}{dx}+ a \sin(a\chi)[\nu+\lambda \cos(\phi)]=0.
\end{equation}
After some algebraic manipulations, the potential can be
cast to the following form:
\begin{eqnarray}
U_1(\phi,\chi)&=&\frac{1}{8}[\lambda^2(1+a^2)+
2(\mu^2+\nu^2 a^2)]+\nonumber \\
& + &\frac{1}{8}[\lambda^2(a^2-1)-2 \mu^2]\cos(2\phi)+\nonumber \\
& + &\frac{1}{8}[\lambda^2(1-a^2)-2 \nu^2 a^2]\cos(2a\chi)+\nonumber \\
& + &\frac{1}{2}\lambda\nu a^2 \cos(\phi)+
\frac{1}{2}\lambda\mu \cos(a \chi)-\nonumber \\
& - &\frac{1}{2}\lambda\mu \cos(2\phi)\cos(a\chi)-
\frac{1}{2}\lambda\nu a^2\cos(\phi)\cos(2a\chi)-\nonumber \\
& - &\frac{1}{8}\lambda^2(1+a^2)\cos(2\phi)\cos(2a\chi).
\end{eqnarray}

Here we notice that in the above general function $(\ref{eq:permod2})$ 
the two first terms correspond to decoupled self-interacting 
scalar field. The self-interacting  fields have different amplitudes, 
controlled by the parameters $\mu$ and $\nu$, and different
periodicities, the difference being governed by the parameter $a$. 
If we set $\lambda =0$ in $(\ref{eq:permod2})$ the two first-order equations
$(\ref{eq:mod2a})$ and $(\ref{eq:mod2b})$ become
\begin{equation}
\frac{d\phi}{dx}+ \mu \sin(\phi)=0,
\end{equation}
and
\begin{equation}
\frac{d\chi}{dx} + \nu a \sin(a\chi)=0.
\end{equation}
These equations can be integrated to give
\begin{equation}
\phi(x)=2\arctan e^{- \mu x},
\end{equation}
and
\begin{equation}
\chi(x)=\frac{2}{a}\arctan e^{- \nu a^2 x},
\end{equation}
which correspond to sine-Gordon systems \cite{raj82}.

The third term in $(\ref{eq:permod2})$ represents interactions between 
the two scalar fields, which is controlled by the parameter $\lambda$. 
The general system contains several parameters, and presents the discrete
$Z_2$ symmetry, and for $a=1$ and $\mu=\nu$ this symmetry becomes the $Z_4$
symmetry. This last case seems to be interesting, since it describes two
identical sine-Gordon systems (represented by the $\phi$ and $\chi$ fields)
interacting with each other. Another interesting case is obtained by setting
$\mu=\nu=0 $ in Eq.~{(\ref{eq:permod2})}, and this will be investigated
in the next subsection. 

If we set $a=1$ and $\mu=\nu$ the potential becomes
\begin{eqnarray}
U_1(\phi,\chi)&=&\frac{1}{4}[\lambda^2+
2\mu^2]-
\nonumber \\
& - &\frac{1}{4}\mu^2[\cos(2\phi)+\cos(2\chi)]+
\nonumber \\
& + &\frac{1}{2}\lambda\mu [\cos(\phi)+ \cos(\chi)-
\nonumber \\
& - &\frac{1}{2}\lambda\mu [\cos(2\phi)\cos(\chi)+
\cos(\phi)\cos(2\chi)-
\nonumber \\
& - &\frac{1}{4}\lambda^2\cos(2\phi)\cos(2\chi).
\end{eqnarray}
Furthermore, the first-order equations get to the form
\begin{equation}
\frac{d\phi}{dx}+ \sin(\phi)[\mu+\lambda \cos(\chi)]=0,
\end{equation}
and
\begin{equation}
\frac{d\chi}{dx}+ \sin(\chi)[\mu+\lambda \cos(\phi)]=0.
\end{equation}

This system presents an infinity set of singular points.
For $|\lambda| < |\mu|$ the singular points are
at $((n\pi,m\pi)$, and alternate between
stable and unstable points. For $|\lambda|>|\mu|$
there are more singular points, and they are located at 
$(\arccos(-\mu/\lambda),\arccos(\mu/\lambda))$.  At the particular 
value $|\lambda|=|\mu|$ one gets to the picture that the system 
presents only stable or unstable points at the center of square 
cells with the four sides being continuous lines of singularities, 
as depicted in Fig. [1].
The following picture then emerges: for
$|\lambda| < |\mu|$ $(|\lambda|>|\mu|)$ the intra (inter)
chain binding is stronger than the inter (intra) one; 
for $|\lambda|= |\mu|$ we have intra and inter chain binding 
equally stronge.

\bigskip

\begin{eqnarray}
\put(-35,0){\line(10,0){70}}
\put(-35,70){\line(10,0){70}}
\put(-35,0){\line(0,10){70}}
\put(35,0){\line(0,10){70}}
\put(0,35){\circle*{3}}
\end{eqnarray}

\begin{center}
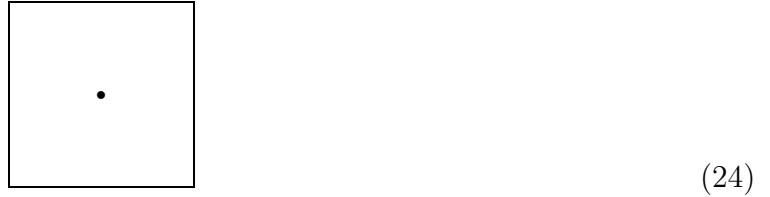
Fig. [1]. A cell of singular points in $(\phi,\chi)$ plane for 
$H_2(\phi,\chi)$\\
when $\mu=\nu=\lambda$ and $a=1$.
\end{center}

\bigskip

We focus our attention at the specific case $|\mu|=|\lambda|$.
To identify topological sectors in this system, we now use the topological 
current introduced at the very end of the former Sect.~{\ref{sec:system}}.
Here we verify that there is only one topological sector, with topological
charge $Q_T=4|\lambda|$. This topological sector represents any of the
infinity possibilities of connecting the central point of the square cell
to a point at its border, by a straight line. See Fig.~[1]. A particularly
simple soliton solution is obtained by setting $\chi=2n\pi$ and
$\lambda=\mu$. In this  case we get, from the above first-order equations,
\begin{equation}
\frac{d\phi}{dx}+ 2\lambda \sin(\phi)=0,
\end{equation}
which is the equation one gets in the sine-Gordon system, and this was already
solved.

\subsection{Polymeric Chain Number Two}

As a second periodic system, we choose $H(\phi,\chi)$
in the following form

\begin{equation}
\label{eq:permod1}
H_2(\phi,\chi)=\lambda\, \cos(\phi)\,\cos(a\chi).
\end{equation}
This system is a particular case of the former system, and we treat it
separately because it will be used in the next section. Here we also note that
other similar forms are obtained by just shifting $\phi\to\phi+\pi/2$ and/or
$\chi\to\chi+\pi/2a$. For instance, the function
$H'_2=\lambda \sin(\phi)\,\sin(a\chi)$ can be considered as an
almost trivial case of the investigation we are now doing,
using the above $H_2$.  

In the present case, however, the first-order equations become
\begin{equation}
\label{eq:mod1a}
\frac{d\phi}{dx}+ \lambda \sin(\phi) \cos(a\chi)=0,
\end{equation}
and
\begin{equation}
\label{eq:mod1b}
\frac{d\chi}{dx}+ \lambda a \cos(\phi) \sin(a\chi)=0.
\end{equation}
The potential can be cast to the form
\begin{eqnarray}
U_2(\phi,\chi)&=&\frac{1}{8}\lambda^2[(1+a^2)
+(a^2 -1)\cos(2\phi)+\nonumber \\
& + &(1-a^2)\cos(2a\chi)-(1+a^2)\cos(2\phi)\cos(2a\chi)].
\end{eqnarray}
Note that the system presents $Z_2$ symmetry, which 
becomes the $Z_4$ symmetry for $a=1$.

This system presents an infinity set of singular points in the 
$(\phi,\chi)$ plane. The singular points can be identified by:
For $n, m =0, \pm 1, \pm 2, ...$, points at
$(2n\pi, 2m\pi/a)$ and at $((2n+1)\pi,(2m+1)\pi/a)$ alternate 
between unstable and stable respectively, and the points
$((2n+1)\pi/2,(2m+1)\pi/2a)$ are all saddle points. See Fig.~[2].

\bigskip

\begin{eqnarray}
\qbezier(-35,0)(-25,0)(-17,17)
\qbezier(-17,17)(-10,35)(0,35)
\put(-35,0){\line(10,0){70}}
\put(0,35){\circle*{3}}
\put(-35,0){\circle*{3}}
\put(35,0){\circle*{3}}
\put(35,70){\circle*{3}}
\put(-35,70){\circle*{3}}
\end{eqnarray}

\begin{center}
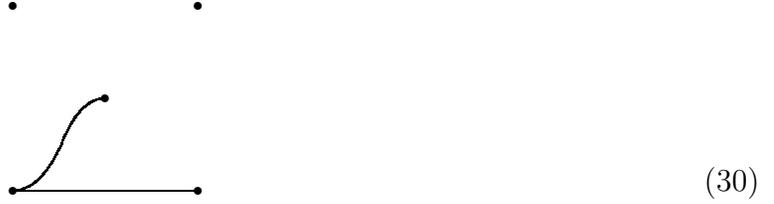
Fig. [2]. A cell of singular points in $(\phi,\chi)$ plane for
$H_1(\phi,\chi)$.\\
Thin lines represent topological sectors.
\end{center}

\bigskip

To identify topological sectors in this system, we now use the 
topological current introduced at the very end of the former 
Sect.~{\ref{sec:system}}. Here we use $(\ref{eq:permod1})$ to 
verify that there are two distinct topological sectors: The first 
one connects adjacent unstable-stable points, which we name the 
paralell sector, since it joins points by a straight line paralell
to the $\phi$ or $\chi$ axis, with the corresponding   
topological charge given by $Q_T^p=2|\lambda|$; the second sector
connects adjacent saddle-stable or unstable-saddle points, which 
we name the transversal sector, in which we have $Q_T^t=|\lambda|$.
Here we notice that the topological  charge (and so the energy) does not
depend on the parameter $a$; also, since $Q^p_T=2Q^t_T$, we see that energy
considerations favor the presence of solitons at the transversal sector.

To find explicit soliton solutions, let us first investigate
the paralell sector. In this case one sets $\chi=2n\pi/a$, and so
the above set of first-order equations $(\ref{eq:mod1a})$ and 
$(\ref{eq:mod1b})$ reduces to the single equation
\begin{equation}
\label{eq:sGordon}
\frac{d\phi}{dx}+ (-1)^n \lambda \sin(\phi)=0.
\end{equation}
This is the sine-Gordon equation that was already studied in the former
subsection.

To investigate the transversal sector, we consider the simplest case:
$a=1$. Here,  we can use $\phi=\chi+2n\pi$, and  the first-order
equations $(\ref{eq:mod1a})$ and $(\ref{eq:mod1b})$ reduce to
\begin{equation}
\frac{d\phi}{dx}+ \frac{1}{2} (-1)^n \lambda \sin(2\phi)=0.\label{28}
\end{equation}
See the former subsection for more details. It is interesting to point out
that the transversal sector is connected by a straight line if and only
if $a=1$. We have being unable to find any explicit analytical solution in 
this transversal sector for $a\ne 1$.

\section{The Deoxyribonucleic Acid}
\label{sec:dna}

In this section we shall consider deoxyribonucleic acid (DNA) as a
system of the form we have been treating in the former sections. Before
doing this, however, we recall that the DNA model proposed by Watson
and Crick is formed by two polynucleotide strands that form the double
helix, which are held together by hydrogen bonds connecting the
adenine-thymine (A-T, with two hydrogen bonds) and guanine-cytosine
(G-C, with three hydrogen bonds) bases. Two-field mechanical models for
DNA consider each one of the two field describing bases in each one of
the two strands, with the two strands being formed by harmonic coupling
responding to the stacking of the DNA chain. Since the intrastrand distance
between bases is at least twice as larger as the corresponding interstrand
distance, such a mechanical view appears to be interesting, although one
still has to account for interations between interstrand pair of bases
governed by hydrogen bonds.

This mechanical picture for DNA is usual, and the complete model in general
differs only on the way one decribes interactions between interstrand pair
of bases, as interestingly considered in \cite{yom83,zha87,pbi89}, and in
references therein. Since we are here interested in finding soliton
solutions, we shall now follow the line of reasoning presented in \cite{zha87}
to model hydrogen bonds in conjugated pair of bases. Before going on,
let us recall that the above mechanical picture of DNA neglects
inhomogeneities due to the base sequence in each strand, and asymetries of
the two strands. Within this context, although the
continuum-limit approximation appears to be somehow severe for DNA
\cite{tho87}, when one discards inhomogeneities in each strand and
asymmetry of the two strands, we realize that the continuum-limit
approximation apeears to be not so severe anymore.

If we now examine the steps introduced in \cite{zha87} to get to the
equation of motion, we recognize that the harmonic interaction present in
each one of the two strands give rise, in the continuum approximation,
to the second-order derivative along the DNA chain. Because of this,
in the potencial there considered, we are just left with several terms,
but all of them related to interactions between conjugate pair of bases
that connects the two strands. This is all we need to jump from this
mechanical model to our system of two coupled fields.

Here we consider the fields $\phi$ and $\chi$ as describing bases in each one
of the two strands in DNA. Moreover, since we are dealing with relativistic
systems, we already have space and time coordinates entering with second
order in derivative in the equations of motion. Then, the function
$H=H(\phi,\chi)$  we have already introduced is to be thought of as
generating the potencial to describe interations between conjugate pair of
bases connecting the two strands, in a way such that if we set to zero those
interactions, we should end up with two independent strands, none of them
containing self-interactions. We then recognize that the polymeric chain
number two, already investigated in the former section, is the field system
that appears to be more adequate to map the DNA model introduced in
Ref.~{\cite{zha87}}.

To see this more explicitly, let us consider the function
\begin{equation}
\label{eq:model3}
H_3(\phi,\chi)=\mu \cos\phi \cos\chi +\nu \sin\phi \sin\chi, 
\end{equation}
which represents the addition of two functions $H(\phi,\chi)$ belonging
to the polymeric chain number two already investigated in the former section.
In this case we get the first-order equations
\begin{equation}
\label{eq:foeqm31}
\frac{d\phi}{dx}+\mu \sin\phi \cos\chi - \nu \cos\phi \sin\chi = 0,
\end{equation}
and
\begin{equation}
\label{eq:foeqm32}
\frac{d\chi}{dx} + \mu\cos\phi \sin\chi - \nu \sin\phi \cos\chi = 0.
\end{equation}
This system has an infinity set of singular points, and this is illustrated
in Fig.~[3]. In this Fig.~[3] the central point is stable (unstable), and the
other four points are unstable (stable). Here we see that there is just one
soliton sector, which connects the central point to one of the other four
points. The energy corresponding to this single soliton sector is given by
\begin{equation}
E_M=|\mu-\nu|.
\end{equation} 

\bigskip

\begin{eqnarray}
\qbezier(-35,0)(-25,0)(-17,17)
\qbezier(-17,17)(-10,35)(0,35)
\put(0,35){\circle*{3}}
\put(-35,0){\circle*{3}}
\put(35,0){\circle*{3}}
\put(35,70){\circle*{3}}
\put(-35,70){\circle*{3}}
\end{eqnarray}

\begin{center}
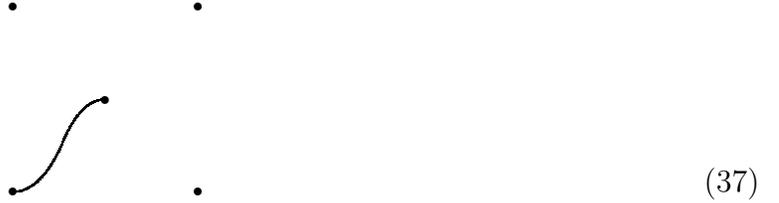
Fig. [3]. A cell of singular points in $(\phi,\chi)$ plane for
$H_3(\phi,\chi)$.\\
The thin line represent the topological sector. 
\end{center}

\bigskip

We deal with the first-order equations $(\ref{eq:foeqm31})$ and
$(\ref{eq:foeqm32})$ to write
\begin{equation}
\frac{d(\phi + \chi)}{dx}=(-\mu+\nu)\sin(\phi + \chi),
\end{equation}
which is solved to give
\begin{equation}
\phi + \chi= 2\arctan e^{(-\mu+\nu)x}\, .
\end{equation}
Furthermore, we can also write
\begin{equation}
\frac{d(\phi - \chi)}{dx}=-(\mu+\nu)\sin(\phi - \chi),
\end{equation}
and now we have
\begin{equation}
\phi - \chi = 2\arctan e^{-(\mu+\nu)x}.
\end{equation}
The pair of solutions are
\begin{equation}
\phi=\arctan e^{(-\mu+\nu)x} + \arctan e^{-(\mu+\nu)x},
\end{equation}
and
\begin{equation}
\chi = \arctan e^{(-\mu+\nu)x} - \arctan e^{-(\mu+\nu)x}.
\end{equation}
It is interesting to see that in this case the potential has the form
\begin{eqnarray}
\label{eq:ourpotential}
U_3(\phi,\chi)&=&\frac{1}{4}\mu^2+ \frac{1}{4}\nu^2- \nonumber\\
& &-\frac{1}{4}\left(\mu^2 +\nu^2\right)\cos2\phi\, \cos2\chi - \nonumber\\
& &-\frac{1}{2}\mu\nu \sin2\phi\, \sin2\chi.
\end{eqnarray}

On the other hand, in \cite{zha87} an interesting  model to map DNA is
considered. There the author uses the following potential
\begin{eqnarray}
U_z(\varphi, \varphi')&=& (-B - 2\beta)\cos \varphi \cos \varphi' +\nonumber\\
& &+(-B+\beta)\sin \varphi \sin \varphi' -\lambda \cos \varphi -
\lambda \cos\varphi',
\end{eqnarray}
where B is the parameter associated with the hydrogen bond energy,
$\beta$ is the parameter associated with the dipole-dipole interaction energy,
and $\lambda$ is the coupling constant associated with the
dipole--induced-dipole interaction energy. These are the parameters that
appear in the DNA model considered in Ref.~{\cite{zha87}}.

Here we see that if one takes, for instance,
\begin{equation}
\mu=\sqrt{2}\sqrt{B + 2 \beta + \sqrt{3\beta(2B+\beta)}},
\end{equation}
and
\begin{equation}
\nu=\sqrt{2}\sqrt{B + 2\beta - \sqrt{3\beta(2B + \beta)}},
\end{equation}
our potential $(\ref{eq:ourpotential})$ exactly reproduces the one
considered in \cite{zha87} in the case $\lambda=0$. This result is very
interesting, and shows that when the dipole--induced-dipole
interaction is absent all the soliton solutions present minimum energy and
are linearly or classically stable.

We now recall that the main motivation presented in \cite{zha87} for studying
soliton solutions in DNA is to perhaps find mechanisms to explain
duplication in DNA. Within this context, if one takes the reasonable point
of view that stable solitons will hardly lead to mechanisms to explain
duplication in DNA, we can very naturally improve the model by just adding
the dipole--induced-dipole interaction term to the potential. In this case we
see that this new potential is not of the form required to belong to the class
of systems already introduced in Sec.~{\ref{sec:system}}, and so we may
perhaps open the possibility of introducing instability for guiding us
toward mechanisms to explain duplication in DNA. To work with this more
realist case, we can proceed as follows: We add the dipole--induced-dipole
interaction term as an extra, infinitesimal contribution to the potential;
then, we follow the procedure introduced in \cite{baz91} to calculate
corrections to the energy of the soliton solutions even when we do not
know the solutions explicitly. Here we recall that the above issue was also
considered in Ref.~{\cite{zha87}}, and there the author introduced another
approach, alternative. On the other hand, when the dipole--induced-dipole
interaction term is turned off we see that the energy $E_M=|\mu-\nu|$ can be
written as $E_M=\sqrt{B}(\Delta_{+} -\Delta_{-})$, with $\Delta_{\pm}$ given by
\begin{equation}
\Delta_{\pm}=\sqrt{2+4\delta \pm 2\sqrt{3\delta(2+\delta)}}\;\; ,
\end{equation}
where we have set $\delta=\beta/B$. Furthermore, as we have already shown, in
this case there is only one topological sector, and the corresponding energy
is given by the above expression. Therefore, this value for the energy may be
seen as a reference value, and can be used as a bound for energy
considerations in investigating duplication in DNA. Furthermore, we could also
consider the polymeric chain number one, and in this case we would introduce
interactions in each one of the two strands, and these self-interactions
would certainly improve the model considered in \cite{zha87} to describe
the DNA chain. Evidently, these issues are out of the scope of the present
paper, and so we postpone to a future work the very specific investigation
concerning the DNA polymeric chain.

\section{Comments and Conclusions}
\label{sec:comments}

In this paper we have investigated a general class of systems of coupled real 
scalar fields. This class of systems is defined by
$$
{\cal L}= \frac{1}{2} \partial_{\alpha}\phi\partial^{\alpha}\phi +
\frac{1}{2} \partial_{\alpha}\chi\partial^{\alpha}\chi -
\frac{1}{2}H^2_{\phi} - \frac{1}{2}H^2_{\chi},
$$
where $H=H(\phi,\chi)$ is a smooth but otherwise arbitrary function of the
two fields $\phi$ and $\chi$. In this case, the second-order equations of
motion corresponding to static field configurations are solved by the
following first-order equations
$$
\frac{d\phi}{dx}=H_{\phi},
$$
and
$$
\frac{d\chi}{dx}=H_{\chi}.
$$

The above set of first-order differential equations can be seen as a dynamical 
system, and so we can take advantage of all the mathematical tools available
to dynamical systems to deal with it. We have also shown that static field
configurations in this class of systems have minimum energy and are
classically or linearly stable, and so the presence of soliton solutions
will certainly play some important role in understanding physical
properties of the system.

Our investigation is done on general grounds, but we have introduced some 
examples of polymeric chains in Sec.~{\ref{sec:periodic}}. There we have
investigated specific systems, described by periodic interactions, with the
motivation of showing how to deal with issues that naturally appear in the
general procedure. As we have shown explicitly, periodic interactions
lead to potentials that can be made periodic too, and the topological
behavior is ease to be identified and can be used to guide us toward finding
soliton solutions.

This fact seem to be important in nonlinear science, since in this case there
are many systems which can be described by periodic interactions among
their degrees of freedom. For instance, as it was shown explicitly,
we have found a model that maps the model used in \cite{zha87}
to describe DNA as a two-field system when the dipole--induced-dipole
interaction term is neglected. This result seems to be interesting since
it informs that without the dipole-induced-dipole interaction term,
the continuum version of the model considered in \cite{zha87} presents
stable soliton solutions, and stable solitons can hardly be used to model
open states in DNA. Within this context, the presence of the
dipole-indiced-dipole interaction seems to be a small correction that
induces instability and so gives rise to a mechanism that may perhaps
lead to open states, thus explaining duplication in DNA. We shall return
to this issue in a future work, in which we deal specifically with solitons
in DNA, owing to explain dupilcation via the presence of open states in
the double helix model of Watson and Crick.

As an ending comment, we would like to add that we can perhaps find other
applications for the models investigated in the present paper. For instance,
we believe that the polymeric chains we have introduced in this work
can be used to map other systems, in particular the model introduced
in \cite{zco94} to describe topological solitons in polyethylene.
These and other related issues are presently under consideration.

\acknowledgments
We would like to thank M. M. Santos for interesting discussions. DB, MTT,
and EV also thank Conselho Nacional de Desenvolvimento Cient\'\i fico
e Tecnol\'ogico, CNPq, Brazil: DB and MTT for partial support and EV
for a fellowship.

\end{document}